\documentclass[11pt]{article}

\usepackage{amssymb,amsthm,amsmath}
\usepackage{algorithm}
\usepackage{algpseudocode}

\usepackage[small]{titlesec}

\makeatletter
 \renewcommand\paragraph{\@startsection{paragraph}{4}{\z@}%
 {1.5ex \@plus1ex \@minus.2ex}%
{-1em}%
{\normalfont\normalsize\bfseries}}
 
\makeatother

\usepackage{subfig}
\usepackage[usenames,dvipsnames]{xcolor}
\usepackage[colorlinks,citecolor=blue,linkcolor=BrickRed]{hyperref}\usepackage[colorlinks,citecolor=blue,linkcolor=BrickRed]{hyperref}
\usepackage{fullpage}
\usepackage{graphicx}
\usepackage{enumitem}
\usepackage{thmtools,thm-restate}
\usepackage{wrapfig}
\usepackage{comment}
\usepackage{cleveref}

\newcommand{\poly}{\ensuremath{\mathsf{poly}}}
\newcommand{\E}{\mathbb{E}}

\newcommand{\tr}{\mathsf{Tr}}
\newcommand{\diag}{\mathsf{diag}}
\newcommand{\op}{\mathsf{op}}

\newcommand{\p}{\mathbb{P}}

\newcommand{\R}{\mathbb{R}}
\newcommand{\Real}{\mathbb{R}}

\newcommand{\Cov}{\mathsf{Cov}}
\renewcommand{\vec}{\mathsf{vec}}
\newcommand{\spn}{\mathsf{span}}
\newcommand{\rank}{\mathsf{rank}}
\newcommand{\adv}{\mathsf{adv}}
\newcommand{\ignore}[1]{{}}

\newtheorem{theorem}{Theorem}[section]

\newtheorem{conjecture}[theorem]{Conjecture}
\newtheorem{lemma}[theorem]{Lemma}

\newtheorem{remark}[theorem]{Remark}

\allowdisplaybreaks

\newenvironment{proofof}[1]{\smallskip\noindent{\bf Proof of #1.}}%
        {\hspace*{\fill}$\Box$\par}

\title{Resolving Matrix Spencer Conjecture Up to Poly-logarithmic Rank}
\author{Nikhil Bansal\thanks{University of Michigan, Ann Arbor. \texttt{bansal@gmail.com}.} \and Haotian Jiang\thanks{University of Washington, Seattle. \texttt{jhtdavid@cs.washington.edu}}
\and
Raghu Meka\thanks{University of California, Los Angeles. \texttt{raghum@cs.ucla.edu}}
}
\date{}

\begin{document}
 
\begin{titlepage}

 \maketitle
  
  \begin{abstract}
  We give a simple proof of the matrix Spencer conjecture up to poly-logarithmic rank: given symmetric $d \times d$ matrices $A_1,\ldots,A_n$ each with $\|A_i\|_{\mathsf{op}} \leq 1$ and rank at most $n/\log^3 n$, one can efficiently find $\pm 1$ signs $x_1,\ldots,x_n$ such that their signed sum has spectral norm $\|\sum_{i=1}^n x_i A_i\|_{\mathsf{op}} = O(\sqrt{n})$. This result also implies a $\log n - \Omega( \log \log n)$ qubit lower bound for quantum random access codes encoding $n$ classical bits with advantage $\gg 1/\sqrt{n}$.

Our proof uses the recent refinement of the non-commutative Khintchine inequality in [Bandeira, Boedihardjo, van Handel, 2022] for random matrices with correlated Gaussian entries.

  \end{abstract}
  \thispagestyle{empty}
  
 \clearpage
%\setcounter{tocdepth}{2}
%\vspace{-0.3cm}
%{\small
%\tableofcontents
%}
\end{titlepage}

\section{Introduction}
We study discrepancy minimization in the matrix setting. Let us start with the classical discrepancy setting where given vectors $a_1,\ldots,a_n \in \Real^d$ satisfying $\|a_i\|_\infty \leq 1$ for all $i\in [n]$, and the goal is to find signs $x_1,\ldots,x_n \in \{\pm 1\}$ to minimize the discrepancy $\|\sum_{i=1}^n x_i a_i\|_\infty$.
In a seminal result, Spencer \cite{S85} showed that the $O(\sqrt{n \log d})$ bound obtained by a random coloring is not tight and showed the following bound, which is also the best possible in general.
\begin{theorem}[Spencer \cite{S85}]\label{th:spencer}
Given vectors $a_1,\ldots,a_n \in \mathbb{R}^d$ each satisfying $\|a_i\|_\infty \leq 1$, there exist signs $x \in \{\pm 1\}^n$ such that $\|\sum_{i=1}^n x_i v_i\|_\infty = O(\sqrt{n} \cdot \max\{1, \sqrt{\log (d/n)} \})$.
\end{theorem}
In particular for $d=O(n)$, this gives an $O(\sqrt{n})$ bound, in contrast to the $O(\sqrt{n\log n})$ bound for random coloring obtained by applying Chernoff and union bounds. 
%The bound in Theorem \ref{} is also the best possible in general.

To prove this result, Spencer developed the powerful partial-coloring method via the entropy method, building on the previous work of Beck \cite{B81}. Another approach to prove Theorem \ref{th:spencer} based on convex geometry was developed independently by Gluskin \cite{Gluskin89}. 
%These approaches have been highly influential in discrepancy theory and its applications \cite{}.
While these original arguments used the pigeonhole principle and were non-algorithmic, in recent years, there has been a rich line of work \cite{B10, BS13, LM15, Rot17, LRR17, ES18, RR20} on their algorithmic versions.

%using techniques from probability, optimization, and convex geometry. This has also led to various new results in areas such as approximation algorithms, differential privacy, machine learning, and statistics. See e.g.~\cite{}. [Rot13, NTZ13, BCKL14, BN17].

\paragraph{Matrix Spencer Setting.}
A natural generalization of Spencer's problem to matrices is the following. Let $A_1,\ldots,A_n \in \Real^{d\times d}$ be symmetric matrices with maximum singular value, or operator norm, $\|A_i\|_{\op} \leq 1$.
Find  a coloring $x \in \{\pm 1\}^n$ that minimizes $\|\sum_{i=1}^n x_i A_i\|_{\op}$. In particular, Spencer's result corresponds to the case when all the $A_i=\diag(a_i)$ are diagonal.

%There has been a long line of work  
As in the vector case, for a random coloring $x \in \{\pm 1\}^n$, the non-commutative Khintchine inequality of Lust-Piquard and Pisier  \cite{LP86,LPP91,Pis03}, or the matrix Chernoff bound \cite{Oliviera, TroppBook}, give that 
\begin{equation}\label{eq:matchernoff}
\E \Big[ \big\|\sum_i x_iA_i \big\|_{\op} \Big] =O\Big( \sqrt{\log d} \cdot \Big\|\sum_i A_i^2 \Big\|^{1/2}_{\op}\Big).
\end{equation}
This implies a bound of $O(\sqrt{n \log d})$ on the matrix discrepancy. This inequality also holds when one picks $x \in \mathbb{R}^n$ to be standard Gaussians, which will play an important role in our results. 

Matrix concentration bounds are powerful and widely used tools in mathematics and computer science, and it is natural to ask when one can beat them.
In particular, whether the following natural analog of Spencer's result for matrices holds is a tantalizing open question.
\begin{conjecture}[Matrix Spencer Conjecture \cite{Z12,M14}] \label{conj:matrix_spencer}
Given $d \times d$ symmetric matrices $A_1, \dots, A_n \in \Real^{d\times d}$ with $\| A_i \|_\op \leq 1$, there exist signs $x\in \{\pm 1\}^n$ such that 
$\big\| \sum_{i=1}^n x_i A_i \big\|_\op \leq O(\sqrt{n} \cdot \max\{1, \sqrt{\log (d/n)}\})$. 
In particular, the matrix discrepancy is $O(\sqrt{n})$ for $d=n$.
\end{conjecture}
\noindent 
While this conjecture is still open, there has been exciting progress on important special cases.
Recently, Hopkins, Raghavendra and Shetty \cite{HRS22} proved \Cref{conj:matrix_spencer} where each matrix $A_i$ has rank $O(\sqrt{n})$; in a different direction, Levy, Ramadas and Rothvoss \cite{LRR17} and Dadush, Jiang and Reis \cite{DJR22} established \Cref{conj:matrix_spencer} for block-diagonal matrices with block size $h = O(n/d)$. 
Recently, Bansal, Jiang, and Meka \cite{BJM22} gave an approach based on \emph{barrier functions} to achieve a bound that unifies and slightly strengthens the results of \cite{HRS22,DJR22}. %See also \cite{} for a result. 

\subsection{Our Results}
\label{subsec:results}

The main result of this paper is the following theorem. 

\begin{restatable}[Matrix Spencer Up to Poly-logarithmic Rank]{theorem}{AlmostMatrixSpencer}
\label{thm:almost_matrix_spencer}
Given $d \times d$ symmetric matrices $A_1,\ldots,A_n \in \mathbb{R}^{d\times d}$ each with $\|A_i\|_{\op} \leq 1$ and $\|A_i\|_F^2 \leq n/\log^3 n$, there exist signs $x \in  \{\pm 1\}^n$ such that $\|\sum_{i=1}^n x_i A_i\|_{\op} = O(\sqrt{n})$. Moreover, these signs can be computed efficiently.
\end{restatable}

Note that the condition $\|A_i\|_F^2 \leq n/\log^3 n$ is satisfied when each $A_i$ has rank at most $n/\log^3 n$ or in particular if $d \leq n/\log^3 n$. Thus \Cref{thm:almost_matrix_spencer} resolves \Cref{conj:matrix_spencer} up to poly-logarithmic dimension or poly-logarithmic rank. We remark that even when assuming the matrices $A_i$ have small rank (even rank $1$) or small dimension (even $d = \sqrt{n}$), it is known that one cannot hope for a bound better than $\Theta(\sqrt{n})$ \cite{DJR22}. % (e.g. Lemma 6.1 and Lemma 6.2 in \cite{DJR22}). 
 For instance, let $e_1,\ldots,e_n \in \mathbb{R}^n$ be the standard basis vectors and take $A_i = (1/2) (e_1 +  e_i)(e_1 + e_i)^T$. Then, for any $x \in \{\pm 1\}^n$, the first column of $\sum_i x_i A_i$ has norm $\Omega(\sqrt{n})$ so its spectral norm is $\Omega(\sqrt{n})$. This is in sharp contrast to the diagonal case, where an $O(\sqrt{r \log n})$ bound for rank $r$ matrices holds \cite{Ban98}, and a $O(\sqrt{r})$ bound was conjectured \cite{BF81}.

Further, when matrices $A_i$ have dimension $d = \omega(n)$ but $\rank(A_i) \leq n/\log^3 n$, the bound in \Cref{thm:almost_matrix_spencer} is $O(\sqrt{n})$ and is stronger than the $\omega(\sqrt{n})$ bound suggested by \Cref{conj:matrix_spencer}. 

%Not only so, when $d = \omega(n)$ but the matrices $A_i$ each satisfies $\rank(A_i) \leq n/\log^3 n$, the bound in \Cref{thm:almost_matrix_spencer} is still $O(\sqrt{n})$, suggesting that the $\omega(\sqrt{n})$ bound in \Cref{conj:matrix_spencer} is too pessimistic in this case.  
%\raghu{Not sure about pessimistic because of the sqrt{n} rank example, so commented the remark.}
%\haotian{My original point is to sell that having a low rank is stronger than the type of statement in the matrix Spencer conjecture (or maybe that the conjecture itself likely only depends on rank). I reworded this sentence, but if this still sounds confusing, I am fine with removing it.}

A new ingredient in our proof is a recent strengthening of the non-commutative Khintchine inequality %matrix Chernoff bound 
for Gaussian random matrices of the form $\sum_i g_i A_i$ where $g_1,\ldots, g_n$ are independent standard Gaussian random variables due to Bandeira, Boedihardjo, and van Handel \cite{BBvH21}. The central idea is to pick a suitable projection of the random matrix to a subspace so that the bound of \cite{BBvH21} matches $O(\sqrt{n})$. We defer the details to the full proof. 

The same proof strategy also implies an improvement over the random coloring bound of $O(\sqrt{n \log d})$ for all $d = o(n)$ by using the result of \cite{Tropp2} together with \cite{BBvH21}\footnote{Specifically, using the bound $\E [\|X\|_{\op}] = O( (\log d)^{1/4} \sigma(X) + (\log d)^{1/2} \sqrt{v(X) \sigma(X)})$ (this follows by combining Corollary 3.6 in \cite{Tropp2} with Proposition 4.6 in \cite{BBvH21}),  instead of the bound given by Theorem \ref{thm:BBvH_matrix_concentration}, in the argument in Section \ref{sec:main} gives the discrepancy bound $o(\sqrt{n \log d})$ for all $d=o(n)$.}. 
%In particular, it implies the bound $\|\sum_{i=1}^n x_i A_i\|_{\op} = O(\sqrt) $

%The proof of the theorem relies on two important ingredients. The first ingredient is a new strengthening of the matrix Chernoff bound for Gaussian random variables due to Bandeira, Boedihardjo, and van Handel \cite{BBvH21}. We then use the known connection between partial colorings and Gaussian volumes of symmetric convex sets (\cite{Rot13}, \cite{Gluskin}, \cite{Gianopolous}). We defer the details to the full proof.

\smallskip
\noindent \textbf{Implications for Quantum Random Access Codes.}
\cite{HRS22} identified a beautiful connection between the matrix Spencer conjecture and quantum random access codes that achieve advantage $C/\sqrt{n}$ for a big enough constant. They use this connection in their proof of the conjecture for matrices of rank $O(\sqrt{n})$. %As an immediate corollary of our result (and the fact that we prove the conjecture when the dimension $d < n/\log^3 n$), we get the following lower bound on quantum random access codes that could be of independent interest. 

Consider the following two-party communication problem: Alice is given a vector $x \in \{\pm 1\}^n$ and Bob an index $i \in [n]$. We are interested in the one-way quantum communication complexity (from Alice to Bob) of computing $x_i$. That is, Alice gets to send a quantum message to Bob and Bob must use this message to compute a guess for $x_i$. For a protocol $\Pi$, let $\adv_{\Pi}(x,i) = \max(0, \p[\Pi(x,i) = x_i] - 1/2)$ be the advantage over random guessing that Alice and Bob have. Note that the randomness is over that of the protocol. 

The seminal works of \cite{ANTV02} showed that for any protocol $\Pi$ with $\E_{x,i}[\adv_{\Pi}(x,i)] = \Omega(1)$, Alice must send $\Omega(n)$ qubits to Bob\footnote{On a related note, if one is not interested in the exact constant, one can obtain an $\Omega(n)$ bound easily from the matrix Chernoff bound in \eqref{eq:matchernoff} (without using any quantum information theory).}. In \cite{HRS22}, the following elegant   connection %\footnote{\raghu{I'm not sure its a good for us to have it here ... it'll be odd}. This connection was independently observed by Sivakanth Gopi without publishing it.}
between \Cref{conj:matrix_spencer} and the above communication problem is made: The conjecture is true if and only if there is some constant $C$ such that any protocol $\Pi$ with $\min_x \E_i[\adv_{\Pi}(x,i)] > C/\sqrt{n}$ must send at least $\log_2 n - O(1)$ qubits from Alice to Bob. 

%Note that the leading factor of $1$ in front the $\log_2 n$ in the above connection is the best possible, even for classical communication. In particular, \Cref{th:spencer} implies that for any constant $C > 0$, there is a classical protocol $\Pi$ with advantage $\min_x \E_i[\adv_\Pi(x,i)] > C/\sqrt{n}$ that uses at most $(1 + o(1)) \log_2 n$ classical bits of communication. 

%\footnote{In the language of \cite{HRS22}, this communication upper bound corresponds to the case where Alice's input is worst-case but Bob's input is uniformly random. A similar upper bound holds when both are worst-case with a slightly more involved protocol.} 

As our main result, \Cref{thm:almost_matrix_spencer}, proves the conjecture for matrices of dimension $n/\log^3 n$, this combined with Claim 1.6 in  \cite{HRS22} immediately imply the following corollary:

\begin{restatable}[QRAC Lower Bound]{corollary}{QRAClb}
\label{cor:QRAClb}
There exists a universal constant $C > 0$ such that the following holds. Any quantum one-way protocol $\Pi$ as above with $\min_x \E_i[\adv_{\Pi}(x,i)] > C/\sqrt{n}$ requires at least $\log_2 n - 3 \log_2 \log_2 n - O(1)$ qubits of communication from Alice to Bob. 
\end{restatable}

Note that the leading constant of $1$ in front of $\log_2 n$ is right for the first time and is the best possible (for sufficiently large constant $C > 0$). Previously, the results of \cite{HRS22,DJR22} imply a lower bound of $(1/2) \log_2 n - O(1)$ on the quantum one-way communication complexity.

Further, a modification of the example in \cite{DJR22} shows that there exists a protocol $\Pi$ such that for all $x \in \{\pm 1\}^n, i \in [n]$, $\adv_{\Pi}(x,i) > c/ \sqrt{n}$ for some constant $c > 0$ and involves at most $(1/2) \log_2 n + O(1)$ qubits of communication. Combined with our lower bound, this shows a somewhat sharp transition in the communication required for protocols as in \Cref{cor:QRAClb}: for some constants $0 < c < C$, achieving an advantage of $C/\sqrt{n}$ requires $\log_2 n - O(\log \log n)$ qubits, whereas one can achieve $c/\sqrt{n}$ advantage with $(1/2) \log_2 n + O(1)$ qubits. Interestingly, the transition is a quantum phenomenon and is absent for classical randomized communication; a tight bound of $\log_2 n + \Theta(\alpha^2)$ bits of communication is known for achieving advantage $\alpha/\sqrt{n}$ for all $\alpha > 0$. 

\subsection{Further Related Works}

\paragraph{Discrepancy Theory.} Discrepancy theory is widely studied and has applications to many other mathematics and computer science areas. We refer readers to the excellent books \cite{C00,M09,CST+14} for a more comprehensive account of the rich
history of discrepancy.  
Recent developments in discrepancy have led to several applications in approximation algorithms, differential privacy, fair allocation, experimental design, and more \cite{MN12, Rot13, NTZ13, BCKL14, Bansal19,JKS19,HSSZ19,BJSS20,BRS22}.

\paragraph{Matrix Discrepancy and Non-Commutativity Random Matrix Theory.} Many natural problems in the study of spectra of matrices can be viewed as questions about matrix discrepancy, e.g., graph sparsification \cite{BSS12, RR20a}, the Kadison-Singer problem \cite{MSS15} and its generalization \cite{KLS20}, and the design of quantum random access codes \cite{ANTV02,HRS22}. %\haotian{Do we need to update this?} 

Matrix discrepancy is also closely related to non-commutative random matrix theory, where the typical value of $\|\sum_i x_i A_i\|_\op$ for a random coloring $x$ has received significant attention. 
The bound of $\E[\|\sum_i x_i A_i\|_\op]\leq O(\sqrt{n \log m})$ by matrix Chernoff \cite{AW02} or matrix Khintchine \cite{LP86,LPP91,Pis03} that is generally tight for commutative matrices, can be often improved in the non-commutative case (e.g. \cite{Ver18,Tropp2,BBvH21} and the references therein). We refer readers to the book \cite{Tao12,Vu14} for a more comprehensive account of random matrix theory.

%\paragraph{Barrier Potential Function Approach.} The inverse polynomial barrier potential function was first used in the seminal work of Batson, Spielman and Srivastava on graph sparsification \cite{BSS12}. Similar potential functions have also been used in the context of discrepancy \cite{BS13,BS20}.  Recently, Bansal, Laddha and Vempala \cite{BLV22} showed how various state-of-the-art results in vector discrepancy can be obtained using this method. We remark that the way the barrier potential function is typically used in discrepancy theory is different from that in \cite{BSS12}. In discrepancy, one typically obtains a fractional update on the coloring, while in \cite{BSS12} a rank-1 update is incurred from adding a single vector. 

\section{Preliminaries}

%\subsection{Notations}
We first recall some basic facts about matrices and describe the notations.
For a square matrix $A \in \Real^{m\times m}$ with entries $a_{ij}$, its trace $\tr(A) = \sum_i a_{ii}$ and Frobenius norm $\|A\|_F = \sqrt{\tr(A^TA)} = (\sum_{ij} a_{ij}^2)^{1/2}$.
If $A$ is symmetric with eigenvalues $\lambda_1,\ldots,\lambda_n$, then we have $\tr(A) = \sum_i \lambda_i$,  $\|A\|_F = (\sum_i \lambda_i^2)^{1/2}$ and its operator norm $\|A\|_{\op} = \max_{\|x\|_2=1} \|Ax\|_2 = \max_i |\lambda_i|$. %For a rectangular matrix $R$ its operator norm $\|R\|_{\op} = \max_{\|x\|_2=1} \|Rx\|_2  = \sqrt{ \|R^T R\|_{\op}}$. 
A symmetric matrix $A$ is positive semidefinite (PSD) if all its eigenvalues $\lambda_i \geq 0$.

For a linear subspace $H \subseteq \mathbb{R}^n$, let $H^\bot$ denote its orthogonal complement. 
For any matrix $A \in \mathbb{R}^{d \times d}$, we let $\vec(A) \in \mathbb{R}^{d^2}$ be the vector formed by the $d^2$ entries of $A$ in a fixed order.  
For a subspace $H$ and convex set $K \subseteq H$, denote $\gamma_H(K)$ the Gaussian measure of $K$ restricted to $H$, i.e. the probability that a standard Gaussian vector on $H$ lies in $K$.

\subsection{Matrix Concentration}
Let $X \in \Real^{d \times d}$ be a symmetric multi-variate Gaussian random matrix (i.e., the entries of $X$ are jointly Gaussian). Equivalently, we can assume that $X$ is of the form $X = \sum_{i=1}^n g_i A_i$ where $g_i$ are independent standard Gaussians and $A_1, \ldots, A_n \in \mathbb{R}^{d \times d}$ are symmetric matrices\footnote{It will be useful to think of the matrix $X$ by itself as a random matrix, and only use the specific representation $\sum_i g_i A_i$ when needed.}.
Note that this representation of $X$ is not unique, and by the rotational invariance of the Gaussians, one also has $X = \sum_j g_j B_j$ where $B_j = \sum_i (v^j)_i A_i$ for any $n\times n$ orthogonal matrix with columns $v^j$.

Let $\sigma(X)^2 = \|\E[X^2]\|_\op = \|\sum_iA_i^2\|_\op$. The fundamental matrix-Chernoff inequality or non-commutative Khintchine inequality implies, among other things, that for $X$ as above, we have
\[\E[\|X\|_\op] = O(\sigma(X) \cdot \sqrt{\log d}) .\]

Note that this bound is tight in general, for instance, if $X$ is a suitable diagonal matrix. Much attention has been given to finding special cases where the $\sqrt{\log d}$ factor in the bound above can be improved. Of particular note is the work of Tropp \cite{Tropp2} where he introduced a specific matrix alignment parameter to capture the non-commutativity of the matrices $A_i$.

%\nikhil{rephrased Tropp's part; the previous version is commented}

%if \emph{independent copies} of $X$ do not commute (or qualitatively if the $A_i$'s do not commute), then the bound can be improved in some sense. 

%While there have been various special cases improving this bound, 
Recently, Bandeira, Boedihardjo, and van Handel made substantial progress in this direction in \cite{BBvH21}. In particular, they related the matrix alignment parameter of Tropp to the following more natural parameter. Let
\begin{align} \label{eq:cov_X}
\Cov(X) := \E[\vec(X) \vec(X)^\top] = \E \Big[\sum_{i=1}^n \vec(A_i) \vec(A_i)^\top \Big]
\end{align} be the $d^2 \times d^2$ covariance matrix of its $d^2$ scalar entries and define 
\[ v(x)^2 := \|\Cov(X)\|_\op.\]
Bandeira, Boedihardjo, and van Handel \cite{BBvH21} showed the following refinement of the non-commutative Khintchine inequality of Lust-Piquard and Pisier \cite{LP86,LPP91,Pis03}. 
\begin{theorem}[\cite{BBvH21}, Theorem 1.2]
\label{thm:BBvH_matrix_concentration}
Given symmetric matrices $A_1,\cdots, A_n \in \mathbb{R}^{d \times d}$, let $X = \sum_{i=1}^n g_i A_i$ where $g_i$ are i.i.d. standard Gaussians. 
Then 
\[
\E[\|X\|_\op] \leq C \cdot (\sigma(X) + (\log^{3/4} d) \sigma(X)^{1/2} v(X)^{1/2} ) ,
\]
where $C$ is some universal constant. In particular,   $\E[\|X\|_\op] = O(\sigma(X) + (\log^{3/2} d) v(X))$. 
\end{theorem}
We remark that the bound in \cite{BBvH21} is substantially more potent and, in particular, gives the optimum constant for the $\sigma(X)$ term and even control over the full spectrum of $X$. However, the weaker version above suffices for our purposes.

%We also need the following corollary of the matrix Bernstein inequality (e.g. \cite{TroppBook}):
%\begin{theorem}\label{thm:matrixbernstein}
%Let $Z_1,\ldots,Z_n \in \mathbb{R}^{d \times d}$ be independent random symmetric matrices with $\E[Z_i] = 0$ and $\|Z_i\|_{\op} \leq 1$. Then we have  $\E[\|\sum_{i=1}^n Z_i\|_{\op}] = O(\sqrt{n \log d})$.
%\end{theorem}

\subsection{Partial Colorings in Convex Sets}
The seminal work of Gluskin \cite{Gluskin89} introduced the idea of finding partial colorings via techniques from convex geometry. At the core is the idea that any symmetric convex set $K \subseteq \R^n$ with sufficiently large Gaussian volume must contain a vector from $\{-1,0,1\}^n$ with $\Omega(n)$ non-zero coordinates (i.e., a \emph{good partial coloring}). In particular, Giannopoulos \cite{Giannopoulos} showed that if $\gamma(K) \geq e^{-\delta n}$ for a sufficiently small constant $\delta$, then $K$ must contain a good partial coloring. Rothvoss \cite{Rot13} gave an algorithmic version of Giannopoulos's result and extended it to subspaces with dimension close to $n$. This extension will be useful for our purposes.  
%in the  It suffices to have $\gamma_H(K) \geq e^{-\delta n}$ for a subspace $H \subseteq \R^n$ as long as $H$ has high dimension.
\begin{lemma}[\cite{Rot17}, Lemma 9] 
\label{lem:rothvoss_partial}
Let $\varepsilon \leq 1/60000$ and $\delta:= \frac{3}{2} \varepsilon \log_2(1/\varepsilon)$. Given a subspace $H \subseteq \mathbb{R}^n$ of dimension at least $(1-\delta) n$, a symmetric convex set $K \subseteq H$ with $\gamma_H(K) \geq e^{-\delta n}$ and a point $x_0 \in (-1,1)^n$. There exists a polynomial time algorithm to find a point $x \in (x_0 + K) \cap [-1,1]^n$ so that $|\{i: x_i \in \{\pm 1\}\}| \geq \varepsilon n/2$. 
\end{lemma}

%\newpage

\section{Proof of the Main Result}
\label{sec:main}
Following the standard approach, it suffices to find a partial coloring with $O(n^{1/2})$ discrepancy. We show this below in Section \ref{sec:pcol}, and then show how Theorem \ref{thm:almost_matrix_spencer}  follows from it in Section \ref{sec:full-col}.

\subsection{Main Partial Coloring Lemma}
\label{sec:pcol}
\begin{lemma}[Main Partial Coloring Lemma] \label{lem:main_partial_coloring}
There exist constants $c, c' > 0$ such that the following holds. Given symmetric matrices $A_1,\ldots,A_n \in \Real^{d \times d}$ that satisfy $\|\sum_{i=1}^n A_i^2 \|_{\op} \leq \sigma^2$ and $\sum_{i=1}^n \|A_i\|_F^2 \leq n f^2$ and a point $x_0 \in (-1,1)^n$, there exists a point $x \in [-1,1]^n$ such that
\[ \Big \|\sum_{i=1}^n (x_i - x_{0,i}) A_i \Big\|_{\op} \leq c (\sigma + (\log^{3/4} d) \sqrt{\sigma f}\,),\]
and $|\{i: x_i \in \{\pm 1\}\}|  > c' n$. Moreover, such a point can be found in polynomial time. 
\end{lemma}

The partial coloring upper bound could be changed to the clearer bound of $O(\sigma + (\log d)^{3/2} f)$ without too much loss; but the above is better for our recursion. In particular, note that if $\sigma \leq \sqrt{n}$ and $f^2 \leq n/\log^{3} d$ (which will be true when $\|A_i\|_{\op} \leq 1$, and $\rank(A_i) \leq n/\log^3 d$), we get a partial coloring with a spectral norm bound of $O(\sqrt{n})$.

%\begin{lemma}[Main Partial Coloring Lemma] \label{lem:main_partial_coloring}
%Given symmetric matrices $A_1,\ldots,A_n \in \Real^{d \times d}$ that satisfy $\|\sum_{i=1}^n A_i^2 \|_{\op} \leq n$ and $\sum_{i=1}^n \|A_i\|_F^2 \leq n^2/\log^3 d$, and a point $x_0 \in (-1,1)^n$. There exists a polynomial time algorithm to find a point $x \in  [-1,1]^n$ such that $\|\sum_{i=1}^n (x_i - x_{0,i}) A_i\|_{\op} = O(\sqrt{n})$, and that $|\{i: x_i \in \{\pm 1\}\}| = \Omega(n)$.
%\end{lemma}

The idea behind the proof is as follows. Let $X = \sum_{i=1}^n g_i A_i$ where $g_i$ are i.i.d standard Gaussian random variables. Consider the convex body 
\[K = \Big\{x \in \R^n: \Big\|\sum_{i=1}^n x_i A_i \Big\|_\op \leq c \sigma \Big\} \subseteq \R^n\] for some suitably large constant $c > 0$. If $K$ had Gaussian measure $\gamma(K) \geq \exp(-\Omega(n))$, then we could directly use Rothvoss's partial coloring result \cite{Rot17}. 
%However, this does not hold as it stands.
%\nikhil{Do we know it is false, or we just don't know?}
As $\sigma(X) \leq \sigma$, one may hope that the improved concentration bound in \Cref{thm:BBvH_matrix_concentration} can be used to show such a lower bound on $\gamma(K)$.  However, it is unclear how to do this directly, as we do not have any control on $v(X)$ and it might even be larger than $\sigma(X)$.
%(besides the bound $v(X) \leq \sigma(X)$ that holds always \haotian{I don't think this is true. Say there is only one $A$, then $\sigma(X)^2 = \|A\|^2_\op$ but $v(X)^2 = \|A\|_F^2.$}) 
So our key idea is to work with a suitable slice of the body $K$.

A key observation is that even if $v(X)$ itself is large, the number of large eigenvalues of $\Cov(X)$ must be small as $\tr(\Cov(X)) = \sum_{i=1}^n \|A_i\|_F^2 \leq n f^2$. In particular, if we set $\Delta^2 \geq f^2/\delta$, then the number of \emph{bad} eigenvectors of $\Cov(X)$ with eigenvalue greater than $\Delta^2$ is at most  $\delta n$. The main idea is to restrict the $g_i$'s to lie in a subspace $H \subseteq \R^n$ so that if $y \in H$ is drawn from the standard Gaussian distribution on $H$, the resulting matrix $Y = \sum_i y_i A_i$ is perpendicular to each of the \emph{bad} eigenvectors of $\Cov(X)$. This ensures that $v(Y) \leq \Delta$ and by \Cref{thm:BBvH_matrix_concentration}, $\E[\|Y\|_\op] = O(\sigma + (\log^{3/4} d) \sqrt{\sigma \cdot f})$. Further, as the number of such bad eigenvectors of $\Cov(X)$ is small, we can ensure that $H$ has dimension at least $(1-\delta)n$. We can now apply \Cref{lem:rothvoss_partial} to get the desired partial coloring. We now give the details.

\begin{proofof}{\Cref{lem:main_partial_coloring}}
Let constants $\varepsilon := 1/60000$ and $\delta := \frac{3}{2} \varepsilon \log_2(1/\varepsilon)$ be as in \Cref{lem:rothvoss_partial}. 
We define $X = \sum_{i=1}^n g_i A_i$ where $g_i$ are i.i.d. standard Gaussian random variables. 
Consider the PSD matrix $\Cov(X) \in \mathbb{R}^{d^2 \times d^2}$ defined in \eqref{eq:cov_X}. Note that by assumption,
\[
\tr(\Cov(X)) = \sum_{i=1}^n \|\vec(A_i)\|_2^2 = \sum_{i=1}^n \|A_i\|_F^2 \leq \delta n \Delta^2,
\]
where $\Delta^2 := f^2/\delta$.
This implies that there can be at most $k := \delta n$ eigenvalues of $\Cov(X)$ exceeding $\Delta^2$. 
Let $V_1, \cdots, V_k \in \mathbb{R}^{d \times d}$ be such that $\vec(V_j)$ is the eigenvector for the  $j$th largest eigenvalue of $\Cov(X)$. Define the subspace 
\[H := \Big\{y \in \mathbb{R}^n: \sum_{i=1}^n y_i \cdot \tr(A_i V_j) = 0, \forall j\in [k] \Big\}.\]
Now we sample the standard Gaussian vector $g \in \mathbb{R}^n$ as follows: first sample a standard Gaussian vector $y \in H$, then sample an independent standard Gaussian vector $r \in H^\bot$, and finally let $g = y + r$. 
We define $Y := \sum_{i=1}^n y_i A_i$ and $R := \sum_{i=1}^n r_i A_i$, which implies $X = Y + R$. 
Since $Y$ and $R$ are independent and have zero mean, we immediately have that $\E[X^2] = \E[Y^2] + \E[R^2]$ and therefore $\sigma(Y) \leq \sigma(X) \leq \sigma$ by the assumption in the Lemma. 

We next show that $v(Y) \leq \Delta$. 
As $\tr(Y V_j) = 0$ for any $j \in [k]$, we have $\vec(V_j)^\top \Cov(Y) \vec(V_j) = 0$. As $\Cov(Y)$ is a PSD matrix, $W:=\spn\{\vec(V_1), \cdots, \vec(V_k)\} \subseteq \R^{d^2}$ must be a subspace of the eigenspace corresponding to the eigenvalue $0$ of the matrix $\Cov(Y)$. 
For any vector $v \in \R^{d^2}$ with  $v \perp W$, we thus have that
\[
v^\top \Cov(Y) v \leq v^\top \Cov(X) v \leq \Delta^2,
\]
as $\Cov(X) = \Cov(Y) + \Cov(R)$, and 
the $(k+1)$th eigenvalue of $\Cov(X)$ is at most $\Delta^2$.
This proves that $\|\Cov(Y)\|_\op \leq \Delta^2$, or equivalently $v(Y) \leq \Delta$. 

Now we want to apply \Cref{thm:BBvH_matrix_concentration} to $Y = \sum_{i=1}^n y_i A_i$. A crucial but elementary fact is that \Cref{thm:BBvH_matrix_concentration} holds for any (symmetric) matrix-valued random variable whose entries are jointly Gaussian and the final bound only depends on the overall distribution of the random matrix and not on the specific representation as a sum of independent matrices. Clearly, the matrix $Y$ we have is a multi-variate Gaussian random variable so we can apply their result. 

To be precise, we can justify its validity even though the vector $y$ does not have independent coordinates as follows. Let $v^1,\ldots,v^k \in \mathbb{R}^n$ be an orthonormal basis for $H$. Then, we can write $y = \sum_{j=1}^k h_j v^j$ where $h_i$ are i.i.d standard Gaussian variables. We can now write
$$Y = \sum_{i=1}^n y_i A_i = \sum_{j=1}^k h_j \Big(\sum_{i=1}^n (v^j)_i A_i \Big) = \sum_{j=1}^k h_j B_j,$$
where we define $B_j = \sum_{i=1}^n (v^j)_i A_i$. Thus $Y$ can be written in the form of a Gaussian matrix series in terms of the i.i.d. standard Gaussians $h_i$. 

Thus we can apply \Cref{thm:BBvH_matrix_concentration} to $Y$ and obtain for universal constant $C > 0$,  
\begin{align} \label{eq:Y_norm}
\E[\|Y\|_\op] \leq C \cdot (\sigma(Y) + (\log^{3/4} d) \cdot \sqrt{\sigma(Y) v(Y)}) \leq c (\sigma + (\log ^{3/4} d) \sqrt{\sigma \cdot f}),
\end{align}
for a sufficiently big constant $c > C (1 + 1/\sqrt{\delta})$. 
Let us consider  the convex body 
\[K' := \Big\{x \in H: \Big\|\sum_{i=1}^n x_i A_i \Big\|_\op \leq 2c (\sigma + (\log ^{3/4} d) \sqrt{\sigma \cdot f}) \Big\}.\] 
By Markov's inequality and \eqref{eq:Y_norm}, it follows that $\gamma_H(K') \geq 1/2 \geq e^{-\delta n}$. 
Also note that $\dim(H) \geq (1-\delta) n$ since $H$ is defined by $\delta n$ constraints. 
It then follows from  \Cref{lem:rothvoss_partial} that we can efficiently find a point $x \in (x_0 + K') \cap [-1,1]^n$ such that $|\{i: |x_i| = 1\}| \geq \varepsilon n/2 = \Omega(n)$. 
By the definition of $K$, the guarantee that $x \in x_0 + K'$ translates to 
\[
\Big \|\sum_{i=1}^n (x_i - x_{0,i}) A_i \Big \|_\op \leq 2 c (\sigma + (\log ^{3/4} d) \sqrt{\sigma \cdot f}).\]
This completes the proof of the lemma. 
As $ (\log ^{3/4} d) \sqrt{\sigma \cdot f} \leq (\sigma + f(\log d)^{3/2})/2$, this implies a partial coloring discrepancy bound of at most 
%The same argument would also show a bound of 
$O(\sigma + (\log d)^{3/2} f)$.
%by using the similar version in \Cref{thm:BBvH_matrix_concentration}.
\end{proofof}

\subsection{Proof of Main Theorem}
\label{sec:full-col}
We can now prove  \Cref{thm:almost_matrix_spencer} (restated below) by recursively applying \Cref{lem:main_partial_coloring}. %We restate \Cref{thm:almost_matrix_spencer} below .

\AlmostMatrixSpencer*

\begin{proofof}{\Cref{thm:almost_matrix_spencer}}
Denote $f^2 := n/\log^3 n$. 
First, without loss of generality, we can assume that $d \leq n^2$. Indeed, suppose to the contrary that $d>n^2$. Define $M := \sum_{i=1}^n A_i^2$ and note that $\tr(M) = \sum_{i=1}^n \|A_i\|_F^2 \leq n f^2$. By a change of basis, we may assume without loss of generality that $M$ is diagonal and its diagonal entries are in descending order. Note that $M_{n^2,n^2} \leq \tr(M)/n^2 \leq f^2/n$. 
Define $B_i \in \mathbb{R}^{(d-n^2) \times d}$ the matrix obtained by removing the first $n^2$ rows of $A_i$. We have for any coloring $x \in \{\pm 1\}^n$,
\[
\Big\| \sum_{i=1}^n x_i B_i \Big\|^2_\op = \Big\| \Big(\sum_{i=1}^n x_i B_i \Big)^\top \Big(\sum_{i=1}^n x_i B_i\Big) \Big \|_\op \leq n \cdot \Big\|\sum_{i=1}^n B_i^\top B_i \Big\|_\op  \leq f^2, 
\]
where the inequality follows as $x_i x_j (B_i^T B_j + B_j^TB_i) \preceq (B_i^TB_i + B_j^T B_j)$ for all $i,j$.
Now we let $L_i \in \mathbb{R}^{d \times d}$ be the matrix obtained by
zeroing out the top left $n^2 \times n^2$ block of $A_i$. Since matrices $A_i$ are symmetric, it follows that  for any coloring $x \in \{\pm 1\}^n$, 
\[
\Big \| \sum_{i=1}^n x_i L_i \Big \|_\op \leq 2 \Big \| \sum_{i=1}^n x_i B_i \Big \|_\op \leq 2 f.
\]
This shows that we only need to keep the top left $n^2 \times n^2$ block of each matrix $A_i$ without affecting the discrepancy by more than an additive term of $2f$. We thus assume henceforth that $d \leq n^2$.  

By assumption, % a constant scaling of
the matrices $A_i$ satisfy $\|\sum_{i=1}^n A_i^2\|_\op \leq n$ and $\sum_{i=1}^n \|A_i\|_F^2 \leq n f^2$. Therefore, we can apply \Cref{lem:main_partial_coloring} with $x_0 = 0$ to obtain a partial coloring $x^{(1)} \in [-1,1]^n$ with $\|\sum_{i=1}^n x^{(1)}_i A_i\|_\op = O(\sqrt{n})$ and $|\{i: |x^{(1)}_i|=1\}| = \Omega(n)$. 
Next we let $I_1 := \{i \in [n]: |x^{(1)}_i| < 1\}$, and recursively apply \Cref{lem:main_partial_coloring} to the set of matrices $\{A_i\}_{i \in I_1}$ with point $x^{(1)}|_{I_1}$.   
Continuing this process of recursively applying \Cref{lem:main_partial_coloring} to
the set of coordinates $i$ such that $|x_i| < 1$, the number of such coordinates decreases by a constant factor in each iteration.

Let $x^{(t)} \in [-1,1]^n$ be the resulting vector in the $t$th iteration and let $n_t$ denote the number of coordinates in $x^{(t)}$ that are in $(-1,1)$. Then, we have $n_{t+1} < \lambda n_t$ for some constant $\lambda < 1$ and by using \Cref{lem:main_partial_coloring} with $\sigma \leq \sqrt{n_t}$, we get that the discrepancy increases additively by at most $c (\sqrt{n_t} + (\log^{3/4} d) \cdot f^{1/2} n_t^{1/4})$. Therefore, repeating it for $O(\log n)$ iterations, we get a full coloring with discrepancy at most
\[ c \sum_t (\sqrt{n_t} + (\log^{3/4} d) \cdot f^{1/2} n_t^{1/4}) = O(\sqrt{n}) + O((\log^{3/4} n)\cdot f^{1/2}  n^{1/4}),
\]
where we have used that $d \leq n^2$ and $n_t$'s form a geometrically decreasing series. The theorem now follows since we have chosen $f^2 = n/\log^3 n$. 
\end{proofof}

%\ignore{$c (\sqrt{n_t} + (\log^{3/2} d) f)$ in iteration $t$.Thus, after $T = O(\log \log d)$ iterations, we have $x_T \in [-1,1]^n$ such that
%\begin{itemize}   \item $\|\sum_{i=1}^n x_{T,i}\; A_i\|_{\op} \leq c \Big(\sum_{t=0}^{T-1} \sqrt{n_t} + T (\log^{3/2} d) f \Big) = O(\sqrt{n} + T (\log^{3/2} d) f) = O(\sqrt{n})$ by our assumption on $f$ and as $d \leq n^2$.
%  \item $|\{i: |x_{T,i}| \neq 1\}| < n/\log d$.
%\end{itemize}
%Let $U = \{i \in [n]: |(x^T)_i| < 1\}$ be the remaining set of fractional coordinates. We next \emph{round} the coordinates in $U$ independently at random to $\{\pm 1\}$: for each $j \in U$, let $y_j \in \{\pm 1\}$ be independent random variables so that $\E[y_j] = x_{T,j}$. 
%Then, by \Cref{thm:matrixbernstein} we have
%\[\E\Big[\Big\|\sum_{j \in U} (y_j - x_{T,j}) A_j \Big\|_{\op}\Big] = O(\sqrt{|U| \cdot \log d}) = O(\sqrt{n}).\]
%We now set $x \in \{\pm\}^n$ to be the full coloring with $x_i = x_{T,i}$ for each $i \notin U$, and $x_i = y_i$ for all $i \in U$. The above bounds together imply that 
%\[\Big\|\sum_i x_i A_i \Big\|_{\op} \leq \Big\|\sum_i x_{T,i} A_i \Big\|_{\op} + \Big\|\sum_{j \in U} (y_j - x_{T,j}) A_j \Big\|_{\op} = O(\sqrt{n}),\]
%as desired. This completes the proof of the theorem. }

% \nikhil{To be fixed. (i) Add some more detail on why it suffices to  keep $n^2\times n^2$ blocks of $A_i$. (ii) Choose $f = n/log^3n$ vs $n/\log^3d$ consistently. (iii) Some more detail on how we apply Lemma 3.1 multiple times? }
% \haotian{See if the above looks ok.}

\begin{remark}
One can also use the version of \Cref{lem:rothvoss_partial} without defining the subspace as in the proof above, but this requires assuming $\|A_i\|_F^2 \leq n/\log^4 d$ in Theorem \ref{thm:almost_matrix_spencer}. In particular, by taking $\vec(A'_i)$ to be the eigenvectors of the matrix $\Cov(X) = \sum_{i=1}^n \vec(A_i) \vec(A_i)^\top$ in descending order of eigenvalues $\lambda_1 \geq \cdots \geq \lambda_n$, the random matrix $\sum_{i=1}^n g_i \lambda_i^{1/2} A_i'$ has the same distribution as $\sum_{i=1}^n g_i A_i$. One can then guarantee that $\gamma_n(K) \geq 2^{-O(n)}$ by considering the event that $g_1, \cdots, g_k$ are all $1/\poly(n)$ small for $k = \Theta(n/\log n)$, and applying \Cref{thm:BBvH_matrix_concentration} to control $\|\sum_{i=k + 1}^n g_i \lambda_i^{1/2} A'_i\|_\op$. 
\end{remark}

\medskip
\noindent \textbf{Acknowledgements.} We thank Ramon van Handel, Victor Reis and Thomas Rothvoss for very helpful comments.

\bibliographystyle{alpha}
\bibliography{bib.bib}

\end{document}